\documentclass[aps,prl,twocolumn,superscriptaddress,showpacs]{revtex4}

\usepackage{amsmath,amssymb,latexsym}
\usepackage{graphicx}
\usepackage{txfonts,mathrsfs}
\usepackage{comment}
\DeclareSymbolFont{symbols}{OMS}{cmsy}{m}{n}
\DeclareSymbolFont{largesymbols}{OMX}{cmex}{m}{n}
\newcommand{\bm}[1]{\boldsymbol #1}

\begin{document}

\title{
Nonthermal antiferromagnetic order and nonequilibrium criticality in the Hubbard model
}

\author{Naoto Tsuji}
\affiliation{Department of Physics, University of Fribourg, 1700 Fribourg, Switzerland}
\author{Martin Eckstein}
\affiliation{Max Planck Research Department for Structural Dynamics, University of Hamburg-CFEL, 22607 Hamburg, Germany}
\author{Philipp Werner}
\affiliation{Department of Physics, University of Fribourg, 1700 Fribourg, Switzerland}

\begin{abstract}

We study dynamical phase transitions from antiferromagnetic to paramagnetic states
driven by an interaction quench in the fermionic Hubbard model using
the nonequilibrium dynamical mean-field theory.
We identify two dynamical transition points where the relaxation behavior qualitatively changes:
one corresponds to the thermal phase transition at which the order parameter decays critically slowly in a power law $\propto t^{-1/2}$,
and the other is connected to the existence of nonthermal antiferromagnetic order in systems with effective temperature above the thermal critical temperature.
The frequency of the amplitude mode extrapolates to zero as one approaches the nonthermal (quasi)critical point, and 
thermalization is significantly delayed by the trapping in the nonthermal state.
A slow relaxation of the nonthermal order is followed by a faster thermalization process. 
\end{abstract}


\date{\today}


\maketitle

In many physical systems out of equilibrium, phase transitions occur 
as a real-time process of symmetry breaking or symmetry recovery. 
Examples for such ``dynamical phase transitions'' include 
the evolution of the Universe \cite{Kibble1976}, liquid helium \cite{Zurek1985}, 
and photoinduced phase transition in solids \cite{Schmitt2008,Yusupov2010,Fausti2011}.
The macroscopic aspects are often described by the time-dependent Ginzburg-Landau theory,
where the order parameter is supposed to vary sufficiently slowly in time and space, so that
the system can be considered to be locally close to thermal equilibrium. 
On the other hand, recent experimental developments of time-resolved measurement techniques 
in solids \cite{Cavalieri2007} and cold atoms \cite{BlochDalibardZwerger2008}
allow one to study
dynamical phase transitions very far from equilibrium on the microscopic time scale of 
correlated quantum systems. In these cases, a ``near-equilibrium'' description might not be applicable. 
For instance, it has been recently suggested that superconductivity can be induced above 
the equilibrium critical temperature ($T_c$) by 
coherently exciting certain lattice vibrations,
and that it lasts for a relatively long time (a few tens of ps)
before thermalization occurs \cite{Fausti2011}. This observation is reminiscent of the prethermalization phenomenon
\cite{BergesBorsanyiWetterich2004, MoeckelKehrein2008, EcksteinKollarWerner2009, KollarWolfEckstein2011},
or the dynamics in the presence of 
a nonthermal fixed point in relativistic quantum field theories \cite{BergesRothkoptSchmidt2008}.
A fundamental question that we pose here is if
the existence of
such a nonthermal fixed point in correlated condensed matter systems 
allows symmetry broken states to survive above $T_c$, and how it affects the dynamics. 

An important and still unresolved issue is 
how to characterize a nonequilibrium phase transition and its critical behavior for quantum systems 
\cite{HohenbergHalperin1977, PolkovnikovSenguptaSilvaVengalattore2011}.
Previous studies have in particular focused 
on the dynamics near quantum phase transitions in low dimensional systems 
(e.~g., 
Refs.~\cite{Feldman2005,MitraTakeiBaekMillis2006,KollathLauchliAltman2007,BarmettlerPunkGritsevDemlerAltman2009}).
Higher dimensional systems are usually expected to show a thermal criticality out of equilibrium
since quantum fluctuations are well suppressed. 
\begin{figure}[ht]
\begin{center}
\includegraphics[width=8.5cm]{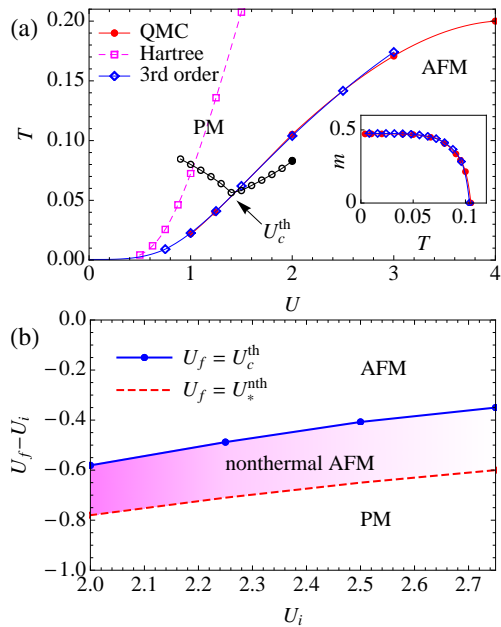}
\caption{
(color online). (a) Equilibrium phase diagram of the Hubbard model in the weak-coupling regime at half filling,
calculated by DMFT with several different impurity solvers.
QMC data are taken from Ref.~\onlinecite{KogaWerner2011}.
Effective temperatures for quenches from a fixed initial state ($U_i=2$, black dot) to various final states (open dots) are shown.
Inset: Staggered magnetization $m$ as a function of $T$ at $U=2$.
(b) Nonequilibrium phase diagram for a quench $U_i\to U_f$ with the fixed initial magnetization, $m(0)=0.4$. 
For $U_f>(<)U_c^{\rm th}$, the system finally thermalizes to an AFM (PM) state. 
A nonthermal AFM order emerges in the colored region. The 
shading indicates the increasing lifetime of the nonthermal AFM state as $U_i$ is reduced.
}
\label{phase diagram}
\end{center}
\end{figure}
In this Letter, we study a dynamical phase transition for a simple microscopic model of correlated materials, namely the Hubbard model.
In equilibrium, the model exhibits a phase transition from paramagnetic (PM) to antiferromagnetic (AFM) order 
[see the phase diagram in Fig.~\ref{phase diagram}(a)]. 
By changing the interaction in time, we cross the phase boundary dynamically.
In particular, we explore the weak-coupling regime of the Hubbard model 
(for the strong-coupling side, see our complementary work \cite{WernerTsujiEckstein2012}).
Contrary to the naive expectation, we find that the nonequilibrium relaxation behavior
can be very different from the thermal one even in the 
large-dimensional
limit. 
A new phenomenon that we demonstrate here is that in addition to the thermal critical point
there exists one more quasicritical point (or sharp crossover) at which some time (energy) scale 
almost diverges (vanishes).
Between these points, the system is trapped in a nonthermal ``ordered'' state [Fig.~\ref{phase diagram}(b)],
where the order parameter stays nonzero even
though 
the effective temperature (which will be defined below) is above $T_c$.

The model Hamiltonian is given by
\begin{align}
H(t)
  &=
    \sum_{\bm k\sigma} \epsilon_{\bm k} c_{\bm k\sigma}^\dagger c_{\bm k\sigma}
    +U(t)\sum_i \left(\hat{n}_{i\uparrow}-\frac{1}{2}\right)\left(\hat{n}_{i\downarrow}-\frac{1}{2}\right),
\nonumber
\end{align}
where $\epsilon_{\bm k}$ is the band dispersion,
$c_{\bm k\sigma}^\dagger$ ($c_{\bm k\sigma}$) is a creation (annihilation) operator of fermions with spin $\sigma$,
$U$ is the (time-dependent) interaction strength, and $\hat{n}_{i\sigma}=c_{i\sigma}^\dagger c_{i\sigma}$.
For convenience, we take a semicircular density of states, $D(\epsilon)=\sqrt{4-(\epsilon/t^\ast)^2}/(2\pi t^\ast)$,
and use $t^\ast$ ($t^\ast{}^{-1}$) as the unit of energy (time). We only show results for the half-filling case.
The initial state is in thermal equilibrium with temperature $T$, which is chosen such that the initial value
of the staggered magnetization 
$m=\langle\,| \hat{n}_{\uparrow}-\hat{n}_{\downarrow}|\,\rangle$ 
is 0.4.
The interaction is changed as $U(t)=U_i+(U_f-U_i)t/t_q$ ($0\le t\le t_q$) with quench time $t_q=8$ fixed.
The interaction quench
can be implemented in cold atom systems with the use of the Feshbach resonance, or by modifying the depth of the lattice potential, 
and has also been proposed to be possible in solids driven by strong electric fields \cite{TsujiOkaWernerAoki2011,TsujiOkaAokiWerner2012}.

The time evolution of the Hubbard model with AFM order is studied 
with the nonequilibrium dynamical mean-field theory (DMFT) \cite{GeorgesKotliarKrauthRozenberg1996,FreericksTurkowskiZlatic2006}.
It becomes exact in the large dimensional limit \cite{MetznerVollhardt1989},
where the self-energy becomes local in space but keeps dynamical correlations. 
When one allows for AFM states in the single-site DMFT,
the self-consistency condition reads $\Lambda_\sigma(t,t')=t^\ast{}^2G_{\bar{\sigma}}(t,t')$ 
\cite{GeorgesKotliarKrauthRozenberg1996,WernerTsujiEckstein2012}
[$\Lambda_\sigma(t,t')$: hybridization function].
Since we are interested in the microscopic dynamics in a single magnetic domain, 
the system is assumed to take a spatially homogeneous configuration.

In order to treat the long-time behavior of symmetry broken states,
we adopt the third-order weak-coupling expansion as an impurity solver,
i.e., expand all the self-energy diagrams, 
including the Hartree term, 
by Weiss Green functions $\mathcal{G}_{0\sigma}(t,t')$ (bare propagators) up to third order in $U$.
Although the bare expansion is not a conserving approximation in the sense of Baym and Kadanoff, 
it turns out to work remarkably well in the weak-coupling regime ($U\lesssim 3$). 
For instance, the total energy is approximately conserved with negligibly small drifts.
By comparison to quantum Monte Carlo (QMC) results \cite{KogaWerner2011}, we confirmed that $T_c$
and $m$ in equilibrium are correctly reproduced [Fig.~\ref{phase diagram}(a)], which is a considerable improvement
from the Hartree approximation [Fig.~\ref{phase diagram}(a)] and
the second-order iterative perturbation theory \cite{GeorgesKotliarKrauthRozenberg1996}.

\begin{figure}[t]
\begin{center}
\includegraphics[width=8.5cm]{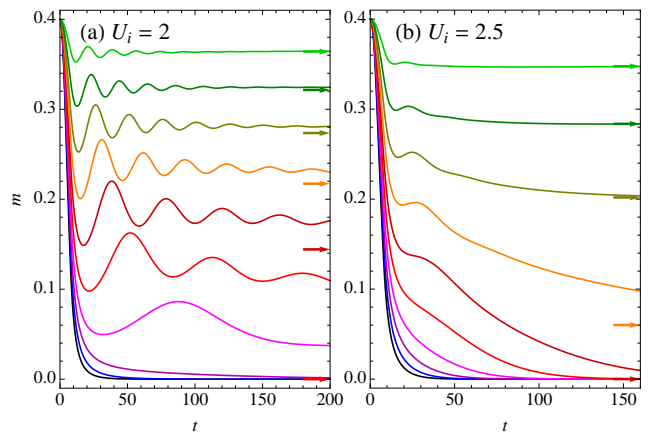}
\caption{(color online). Time evolution of $m$ for quenches (a) 
$U_i=2 \to U_f=1.0, 1.1, \dots,1.9$ (from bottom to top), and
(b) $U_i=2.5 \to U_f=1.5, 1.6, \dots, 2.4$.
The arrows indicate the corresponding thermal values $m_{\rm th}$ reached in the long-time limit.}
\label{order parameter}
\end{center}
\end{figure}
Let us first look at results for quenches  from
$U_i=2$ to various $U_f (<U_i)$.
As shown in Fig.~\ref{order parameter}(a),
$m(t)$ quickly decreases after the quench
due to the reduction of $U$, and starts to oscillate coherently 
(amplitude mode) with a slow drift. 
As $U_f$ decreases below $\sim 1.2$, the oscillation disappears,
and $m$ exponentially decays to zero.
Assuming that  the nonintegrable Hubbard model thermalizes, the
long-time limit of the order parameter is determined by the thermal value $m_{\rm th}$
at some effective temperature  $T_{\rm eff}$. 
Since the total energy is conserved after the quench ($t\ge t_q$) in the isolated system, 
$T_{\rm eff}$ is given by the temperature of the equilibrium system with the same total energy.
The final thermalized states are plotted as open dots in Fig.~\ref{phase diagram}(a).
Since we are considering rather slow changes ($t_q=8$) of $U$, the final states roughly keep track of the constant entropy curve
\cite{WernerParcolletGeorgesHassan2005}.

The evaluated $m_{\rm th}$ are indicated by arrows in Fig.~\ref{order parameter}, and are plotted as a function of $U_f$ in
Fig.~\ref{trapped state}.
One notices that the center of the oscillation of $m$ deviates more and more from $m_{\rm th}$ as $U_f$ is reduced.
Surprisingly, at $U_f=U_c^{\rm th}=1.42$, where $m_{\rm th}$ vanishes $\propto |U_f-U_c^{\rm th}|^\beta$ 
(Fig.~\ref{trapped state})
with the mean-field exponent $\beta=\frac{1}{2}$ (thermal phase transition), 
$m$ still exhibits oscillations around a nonzero value
for a long time. This suggests that the system is effectively trapped in a nonequilibrium quasisteady state, or close to a nonthermal fixed point, 
which allows for a long-lived symmetry broken state
with $T_{\rm eff}$ above $T_c$. In the paramagnetic phase, the system shows prethermalization 
\cite{MoeckelKehrein2008, EcksteinKollarWerner2009, KollarWolfEckstein2011};
i.e., the momentum-integrated quantities such as the double occupancy thermalize faster than
momentum-dependent quantities (e.~g., the momentum distribution). 
Here a new observation is that the order parameter $m$, even though it is momentum integrated,
also stays nonthermal, allowing the symmetry-broken state to survive for 
a long time. This can be attributed to the presence
of ``classical fluctuations'' \cite{BergesRothkoptSchmidt2008} in the Hartree term, which 
is absent in the paramagnetic phase.

\begin{figure}[t]
\begin{center}
\includegraphics[width=8.5cm]{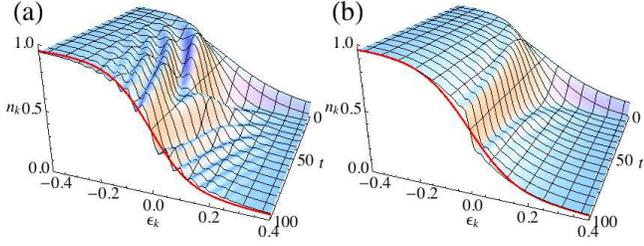}
\caption{(color online). Time evolution of the momentum distribution $n_{\bm k}$ for quenches 
$U_i=2 \to$ (a) $U_f=1.4$ and (b) $U_f=1.2$. The curves at $t=100$ are thermal distributions achieved in the long-time limit.}
\label{momentum distribution}
\end{center}
\end{figure}
To look at the qualitative change of the relaxation behavior around $U_f\sim 1.2$ more closely, 
we calculate the momentum distribution 
$n_{\bm k}(t)\equiv \langle c_{\bm k\sigma}^\dagger(t)c_{\bm k\sigma}(t)\rangle$ \cite{Supplementary}.
In Fig.~\ref{momentum distribution}, one can clearly see the qualitative difference of $n_{\bm k}$
between (a) $U_f=1.4$ and (b) $U_f=1.2$. 
In the former case, waves are continuously generated at high energy,
and cascade down to the lower energy region. They
eventually reach 
the Fermi energy $\epsilon_{\bm k}=0$, and
lead to  
an oscillation of the slope $\partial_\epsilon n$ at $\epsilon_{\bm k}=0$ \cite{Supplementary}.
In the latter case, the wave fronts never arrive at the Fermi energy but accumulate near $\epsilon_{\bm k}=0$,
which results in a steepening slope $\partial_\epsilon n$. This evolution 
is opposite to a heating effect, where
an initially sharp 
momentum distribution is smeared out.
Since the $n_{\bm k}$ in Fig.~\ref{momentum distribution}(b) is very different from a thermal distribution [curve at $t=100$ in Fig.~\ref{momentum distribution}(b)]
the fast relaxation of $m$ for $U_f\le 1.2$ [Fig.~\ref{order parameter}(a)]
is due to dephasing, not thermalization.

To characterize the nonthermal transition observed around $U\sim 1.2$ quantitatively, 
we evaluate the relaxation time $\tau_{\rm deph}$
for the dephasing of $m(t)$ by fitting with $e^{-t/\tau_{\rm deph}}$. As shown in Fig.~\ref{trapped state}(a),
the dephasing critically slows down as 
$\tau_{\rm deph}\propto |U_f-U_\ast^{\rm nth}|^{-1}$
with $U_\ast^{\rm nth}=1.23$ (nonthermal transition point).
At $U_f=U_\ast^{\rm nth}$, $m(t)$ shows a 
power-law decay of $t^{-1/2}$
until thermalization starts to take place around $t\sim 100$.
This indicates that one more quasicritical point with an associated diverging time scale
exists away from the thermal critical point ($U_f=U_c^{\rm th}$). 
Moreover, a sharp kink is observed at $U_f=U_\ast^{\rm nth}$ 
in the plot of the inverse of the steepest slope
$(\partial_\epsilon n)^{-1}=(\max_t\{|\partial_\epsilon n(t)|\})^{-1}$ at $\epsilon_{\bm k}=0$
[Fig.~\ref{trapped state}(a)].
Because a true discontinuity in the momentum distribution function, with $(\partial_\epsilon n)^{-1}=0$, 
would correspond to a power-law decay of the density correlations in space, 
one may thus note that at the nonthermal critical point the system evolves through an almost ``critical  
state'' before thermalization sets in.
We also determined the frequency $\omega_m$
of the amplitude mode of $m$ and the frequency $\omega_{\partial_\epsilon n}$ of the oscillation of $\partial_\epsilon n$ at $\epsilon_{\bm k}=0$
for $U_f>U_\ast^{\rm nth}$ by measuring the peak-to-dip distance
of the oscillations. Note that near the critical point the period of the oscillation exceeds the lifetime ($\sim 100$) of the trapped state,
so that a meaningful measurement is not possible.
However, the 
results in Fig.~\ref{trapped state}(a) indicate that $\omega_m$ and $\omega_{\partial_\epsilon n}$ 
extrapolate to zero
as $\sim |U_f-U_\ast^{\rm nth}|$. 
Based on this fact, we conclude that the amplitude mode is associated with the nonthermal fixed point, not with the thermal phase transition. 
This is not expected in the Ginzburg-Landau picture, where the oscillation disappears
when the curvature of the free energy potential at the origin changes sign at the thermal critical point.
\begin{figure}[t]
\begin{center}
\includegraphics[width=8.5cm]{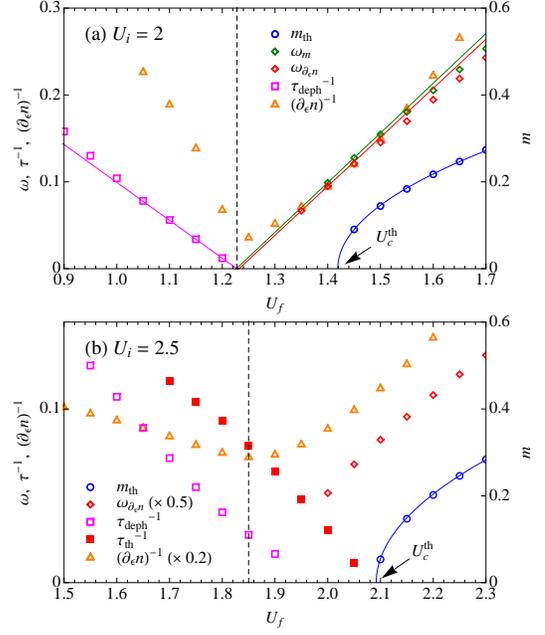}
\caption{(color online). Various quantities 
used 
to characterize the qualitative change of the behavior
around $U_f=U_c^{\rm th}$ and $U_f=U_\ast^{\rm nth}$ 
(dashed lines)
for quenches $U_i\to U_f$. Solid lines are guides for the eye.}
\label{trapped state}
\end{center}
\end{figure}

This quasicritical point (or sharp crossover) becomes an exact critical point in the weak-correlation limit,
where the dynamics is described by the Hartree approximation. As we show in the Supplemental Material,
the Hartree equation is 
mathematically equivalent to 
the time-dependent BCS equation, which is known to be integrable 
with infinitely many conserved quantities \cite{BarankovLevitov2006,YuzbashyanDzero2006}.
There is a strict transition for the motion of the order parameter from 
damped oscillation to overdamped decay
that is both associated with a diverging dephasing time (overdamped decay) and a vanishing of the amplitude-mode frequency. 
What we found here is that the qualitative aspects of the transition are maintained
even in the regime where the Hartree approximation breaks down ($U\gtrsim 0.5$)  
due to quantum corrections from higher-order diagrams. 
In fact, the Hartree equation gives quantitatively very different results in this regime \cite{Supplementary}.

\begin{figure}[t]
\begin{center}
\includegraphics[width=8.5cm]{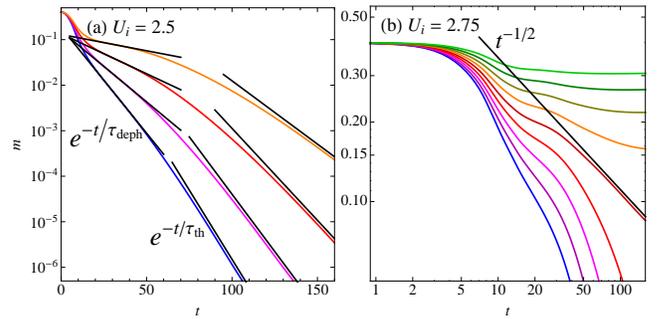}
\caption{(color online). (a) The log plot of $m$ for quenches $U_i=2.5\to U_f=1.6, 1.7, 1.8, 1.9$ from
bottom to top. The straight lines 
show fits of the two exponential relaxations.
(b) The log-log plot of $m$ for quenches $U_i=2.75\to U_f=2.2, 2.25, \dots, 2.6$
from bottom to top.
The straight line shows the slope of a power law decay $\propto t^{-1/2}$.}
\label{power law}
\end{center}
\end{figure}

As one increases $U_i$, the system spends less time near the nonthermal fixed point, and thermalization
occurs earlier. For $U_i=2.5$ [Fig.~\ref{order parameter}(b)], 
coherent amplitude oscillations are not visible 
anymore, and only a bump structure remains
on a short time scale ($t\lesssim 30$) for $U_f> 1.8$. 
In this interaction regime the system does not show a clear signature of a transition, 
but a nonthermal crossover behavior is still seen in various quantities [Fig.~\ref{trapped state}(b)] around $U_f=U_\ast^{\rm nth}\sim 1.85$,
which is estimated from the maximum of $\partial_\epsilon n$.
For $U_f<U_\ast^{\rm nth}$, 
we find that the order parameter $m$ shows a two-step relaxation [Fig.~\ref{power law}(a)]; 
i.e., the short-time and long-time dynamics have different exponential decay rates.
The former is identified to be $\tau_{\rm deph}$, since it is smoothly connected to what we have defined as $\tau_{\rm deph}$ 
in the previous $U_i=2$ case. The latter is related to the thermal phase transition where $m_{\rm th}$ disappears, 
hence denoted by $\tau_{\rm th}$ \cite{limitation}.
The obtained $\tau_{\rm deph}$ and $\tau_{\rm th}$ are shown in Fig.~\ref{trapped state}(b). 
Interestingly, in most cases $\tau_{\rm deph}$ is larger than $\tau_{\rm th}$, that is, the slow dephasing of $m$
is followed by faster thermalization. Furthermore, thermalization is significantly delayed compared to $\tau_{\rm th}$.
At $U_f=1.9$, for example, $\tau_{\rm th}=15.3$ while the delay time of thermalization is $>100$.
This allows the order parameter to survive longer than the thermalization time constant.

Finally, let us examine 
the relaxation around the thermal critical point.
Thermalization critically slows down as one approaches the thermal critical point 
[Fig.~\ref{trapped state}(b)]
with
\begin{align}
\tau_{\rm th}\propto |U_f-U_c^{\rm th}|^{-1},
\label{tau_th}
\end{align}
which, unlike $\tau_{\rm deph}$, remains even when the interaction is increased.
Since the critical behavior around the thermal transition is universal, i.e., does not depend on details of the initial state or the
the ramp protocol, it can be described 
by equilibrium properties. In fact, near a thermal (or quantum) critical point 
the relaxation time is known to behave as $\tau_{\rm th}\sim |U_f-U_c^{\rm th}|^{-z\nu}$ \cite{HohenbergHalperin1977}.
Here $\nu$ is the critical exponent that characterizes the divergence of the correlation length, $\xi\sim |U_f-U_c^{\rm th}|^{-\nu}$,
and $z$ is the dynamical critical exponent. Our result (\ref{tau_th}) is consistent with the mean-field exponents $\nu=\frac{1}{2}$ 
and $z=2$ for nonconserved order parameters \cite{HohenbergHalperin1977}.
Exactly at the thermal critical point ($U_f=U_c^{\rm th}$), 
the correlation time diverges, and 
the order parameter thermalizes in a power law.
In Fig.~\ref{power law}(b),
we show the log-log plot of $m$ around the thermal critical point ($U_c^{\rm th}=2.40$). The curve agrees very well with
\begin{align}
m\propto t^{-1/2}.
\label{power law in m}
\end{align}
This is consistent to the prediction of the dynamical scaling ansatz \cite{HohenbergHalperin1977},
$m\sim t^{-\beta/z\nu}$, with the mean-field exponent $\beta=\frac{1}{2}$.

We summarize our results in a nonequilibrium phase diagram in 
Fig.~\ref{phase diagram}(b). 
The results do not qualitatively change away from half filling \cite{Supplementary} or with different
initial $m$ or $T$. In fact, we numerically confirmed with
the Hartree equation and the nonequilibrium DMFT that the slightly doped ($\lesssim 5\%$) system
can be trapped in a nonthermal ordered state, and that the ``critical'' behavior
at the nonthermal fixed point is the same.
Our findings are applicable not only to antiferromagnetic order but also to superconductivity and charge density wave order
if one translates the repulsive model to an attractive model \cite{Shiba1972}.
An open question of practical importance is how to access this nonthermal fixed point.
While we focused here on interaction quenches, 
the phenomenon is not specific to the particular quench protocol. For example, we have confirmed
that a back-and-forth quench \cite{Supplementary} gives similar nonthermal critical behavior 
with elevated $T_{\rm eff}$,
implying that the overall change of the interaction parameter is not essential.
This universality nature of the phenomenon will open up a possible route 
to experimentally reach the nonthermal fixed point
such as heating the system with laser irradiation. 
Since the order parameter is connected to the energy gap,
the nonthermal order can be monitored with time-resolved optical and photoemission spectroscopies.

We thank H. Aoki, P. Barmettler, J. Berges, and T. Oka for fruitful discussions.
The calculations were carried out on the Brutus cluster at ETH Zurich and on the UniFr cluster.
We acknowledge support from the Swiss National Science Foundation (Grant No. PP0022-118866) and FP7/ERC Starting Grant No. 278023.

\section{Supplemental Material}

\subsection{Hartree approximation}

In this section, we derive the Hartree approximation for the antiferromagnetic phase of the Hubbard model, and show that
it is mathematically equivalent to the integrable time-dependent BCS equation \cite{BarankovLevitov2006, YuzbashyanDzero2006}
at arbitrary filling.

Let us define a set of momentum distribution functions
using the nonequilibrium (lesser) Green function,
\begin{align}
n_{\bm k\sigma}^{ab}(t)
&=(-i)G_{\bm k\sigma}^{ab<}(t,t)
\label{nkabdef}
\\
&=N^{-1}\sum_{i\in a, j\in b}e^{i{\bm k}\cdot({\bm R}_i-{\bm R}_j)}\langle c_{i\sigma}^\dagger(t)c_{j\sigma}(t)\rangle
\end{align}
with $a, b=A, B$ sublattice indices and $N$ the number of sublattice sites. The nonequilibrium
Green function satisfies the 
$2\times 2$
Dyson equation
\begin{align}
&
\begin{pmatrix}
i\overrightarrow{\partial_t}+\mu-\Sigma_\sigma^A & -\epsilon_{\bm k}\\
-\epsilon_{\bm k} & i\overrightarrow{\partial_t}+\mu-\Sigma_\sigma^B
\end{pmatrix}
\ast
\begin{pmatrix}
G_{\bm k\sigma}^{AA} & G_{\bm k\sigma}^{AB}\\
G_{\bm k\sigma}^{BA} & G_{\bm k\sigma}^{BB}
\end{pmatrix}
\nonumber
\\
  &=
    \begin{pmatrix}
    \delta_{\mathcal{C}} & 0\\
    0 & \delta_{\mathcal{C}}
    \end{pmatrix},
\label{2x2 Dyson}
\end{align}
and its conjugate equation
\begin{align}
&
\begin{pmatrix}
G_{\bm k \sigma}^{AA} & G_{\bm k \sigma}^{AB}\\
G_{\bm k \sigma}^{BA} & G_{\bm k \sigma}^{BB}\\
\end{pmatrix}
\ast
\begin{pmatrix}
-i\overleftarrow{\partial_{t'}}+\mu-\Sigma_{\sigma}^A & -\epsilon_{\bm k} \\
-\epsilon_{\bm k} & -i\overleftarrow{\partial_{t'}}+\mu-\Sigma_{\sigma}^B
\end{pmatrix}
\nonumber
\\
  &=
    \begin{pmatrix}
    \delta_{\mathcal{C}} & 0 \\
    0 & \delta_{\mathcal{C}}
    \end{pmatrix},
\label{2x2 Dyson conjugate}
\end{align}
where $\mu$ is the chemical potential, $\Sigma_\sigma^a$ is the local self-energy
on sublattice $a$, $\ast$ denotes a convolution for time arguments, and
$\delta_{\mathcal{C}}$ is the delta function defined on the Keldysh contour $\mathcal{C}$.
In the Hartree approximation, the self-energy is given by
\begin{align}
\Sigma_\sigma^A(t,t')
  &=
    U(t)n_{\bar{\sigma}}^A(t)\delta_{\mathcal{C}}(t,t'),
\\
\Sigma_\sigma^B(t,t')
  &=
    U(t)n_{\bar{\sigma}}^B(t)\delta_{\mathcal{C}}(t,t'),
\end{align}
with $n_\sigma^a(t)=\langle c_{i\sigma}^\dagger(t)c_{i\sigma}(t)\rangle$ ($i\in a=A, B$ sublattice)
the local density. In the presence of AFM order,
the local densities are simply
\begin{align}
n_\sigma^A(t)
&=
\bar{n}+\frac{1}{2}\sigma m(t),
\\
n_\sigma^B(t)
&=
\bar{n}-\frac{1}{2}\sigma m(t),
\end{align}
where $\bar{n}$ is the average density per site and spin.
Using the Dyson equations (\ref{2x2 Dyson}) and (\ref{2x2 Dyson conjugate}) with the Hartree approximation,
we obtain
a closed set of equations of motion for the equal-time lesser Green functions,
\begin{align}
(i\partial_t+i\partial_{t'})G_{\bm k \sigma}^{AA<}(t,t')|_{t' =t}
  &=
    \epsilon_{\bm k}\left[G_{\bm k \sigma}^{BA<}(t,t)-G_{\bm k \sigma}^{AB<}(t,t)\right],
\\
(i\partial_t+i\partial_{t'})G_{\bm k \sigma}^{BB<}(t,t')|_{t'=t}
  &=
    -\epsilon_{\bm k}\left[G_{\bm k \sigma}^{BA<}(t,t)-G_{\bm k \sigma}^{AB<}(t,t)\right],
\\
(i\partial_t+i\partial_{t'})G_{\bm k \sigma}^{BA<}(t,t')|_{t'=t}
  &=
    \epsilon_{\bm k}\left[G_{\bm k \sigma}^{AA<}(t,t)-G_{\bm k \sigma}^{BB<}(t,t)\right]
\nonumber
\\
  &\quad
    -U(t)m(t)\bar{\sigma}G_{\bm k \sigma}^{BA<}(t,t),
\\
(i\partial_t+i\partial_{t'})G_{\bm k \sigma}^{AB<}(t,t')|_{t'=t}
  &=
    -\epsilon_{\bm k}\left[G_{\bm k \sigma}^{AA<}(t,t)-G_{\bm k \sigma}^{BB<}(t,t)\right]
\nonumber
\\
  &\quad
    +U(t)m(t)\bar{\sigma}G_{\bm k \sigma}^{AB<}(t,t).
\end{align}
Equation (\ref{nkabdef}) is used to replace the equal-time Green functions by the corresponding momentum distribution 
functions, with which the equations read
\begin{align}
\partial_t \left[n_{\bm k \sigma}^{AA}(t)+n_{\bm k \sigma}^{BB}(t)\right]
  &=
    0,
\label{nAA+nBB}
\\
\partial_t \left[n_{\bm k \sigma}^{BA}(t)+n_{\bm k \sigma}^{AB}(t)\right]
  &=
    -iU(t)m(t)\sigma\left[n_{\bm k \sigma}^{BA}(t)-n_{\bm k \sigma}^{AB}(t)\right],
\\
i\partial_t \left[n_{\bm k \sigma}^{BA}(t)-n_{\bm k \sigma}^{AB}(t)\right]
  &=
    2\epsilon_{\bm k}\left[n_{\bm k \sigma}^{AA}(t)-n_{\bm k \sigma}^{BB}(t)\right]
\nonumber
\\
  &\quad
    +U(t)m(t)\sigma\left[n_{\bm k \sigma}^{BA}(t)+n_{\bm k \sigma}^{AB}(t)\right],
\\
\partial_t \left[n_{\bm k \sigma}^{AA}(t)-n_{\bm k \sigma}^{BB}(t)\right]
  &=
    -2i\epsilon_{\bm k}\left[n_{\bm k \sigma}^{BA}(t)-n_{\bm k \sigma}^{AB}(t)\right],
\end{align}
To make the expression transparent, we adopt 
a representation analogous to
Anderson's pseudospin for the BCS theory \cite{Anderson1958}, 
\begin{align}
f_{\bm k}^x(t)
  &=
    \frac{1}{2}\sum_\sigma [n_{\bm k\sigma}^{BA}(t)+n_{\bm k\sigma}^{AB}(t)],
\\
f_{\bm k}^y(t)
  &=
    \frac{i}{2}\sum_\sigma \sigma [n_{\bm k\sigma}^{BA}(t)-n_{\bm k\sigma}^{AB}(t)],
\\
f_{\bm k}^z(t)
  &=
    \frac{1}{2}\sum_\sigma \sigma [n_{\bm k\sigma}^{AA}(t)-n_{\bm k\sigma}^{BB}(t)].
\end{align}
Then the Hartree equation can be simply written in the form of a `Bloch equation',
\begin{align}
\partial_t\bm f_{\bm k}(t)
  &=
    {\bm b}_{\bm k}(t) \times {\bm f}_{\bm k}(t),
\label{Hartree equation}
\end{align}
with the pseudospin
\begin{align}
\bm f_{\bm k}
  &=
    (f_{\bm k}^x,f_{\bm k}^y,f_{\bm k}^z)
\end{align}
and an effective magnetic field
\begin{align}
\bm b_{\bm k}(t)
  &=
    (-2\epsilon_{\bm k}, 0, U(t)m(t)).
\end{align}
The order parameter is self-consistently determined by 
\begin{align}
m(t)
  &=
    \sum_{\bm k}f_{\bm k}^z(t).
\end{align}
It turns out that
the equation (\ref{Hartree equation}) is mathematically equivalent to the time-dependent BCS equation \cite{BarankovLevitov2006, YuzbashyanDzero2006}
if one appropriately translates the order parameter from the repulsive model to an attractive model
\cite{Shiba1972}.
The equation is known to be integrable, and has infinitely many conserved quantities. 
For example, $n_{\bm k\sigma}^{AA}+n_{\bm k\sigma}^{BB}$ 
[Eq.~(\ref{nAA+nBB})]
and the `length of the pseudospin' 
\begin{align}
|\bm f_{\bm k}|^2
  &\equiv 
    (f_{\bm k}^x)^2+(f_{\bm k}^y)^2+(f_{\bm k}^z)^2
\label{length of pseudospin}
\end{align}
are conserved for each $\bm k$.

\begin{figure}[t]
\begin{center}
\includegraphics[width=8.5cm]{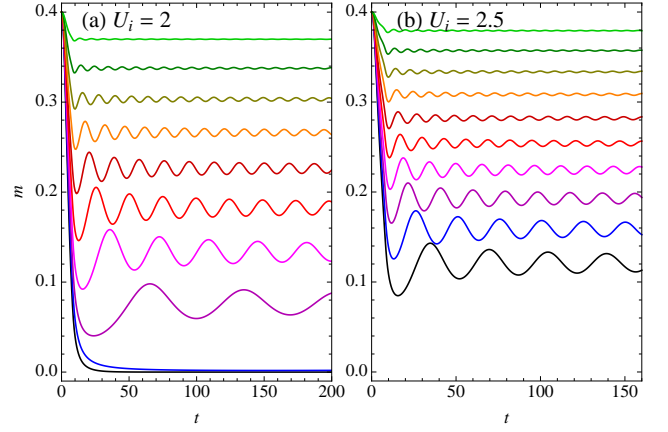}
\caption{Time evolution of $m$ calculated with the Hartree approximation for quenches
(a) $U_i=2 \to U_f=1.0, 1.1, \dots, 1.9$ (from bottom to top) and (b) $U_i=2.5 \to U_f=1.5, 1.6, \dots, 2.4$ (from bottom to top).
The color codes are the same as Fig.~2 in the main text. Note that the temperatures of the initial states are
chosen to be different values from those of Fig.~2 in the main text to fix $m(0)=0.4$.}
\label{order parameter Hartree}
\end{center}
\end{figure}
Figure \ref{order parameter Hartree} shows the results obtained from the Hartree approximation for the order parameter $m$.
We choose the different temperatures of the initial states from those of Fig.~2 in the main text
such that the initial value of the order parameter is the same as in the main text ($m(0)=0.4$).
For $U_i=2$ [Fig.~\ref{order parameter Hartree}(a)],
the overall tendency of the behavior is qualitatively similar to what we have observed
with the nonequilibrium DMFT calculation in Fig.~2(a) of the main text, apart from the fact that
the system never thermalizes within the Hartree calculation,
such that the nonthermal transition  becomes infinitely sharp.
After the quench, $m$ rapidly decreases, and is suddenly trapped
in a nonthermal value with
coherent oscillation of the amplitude. 
Between $U_f=1.1$ and $1.2$, the center of the oscillations gradually approaches zero,
and the behavior of $m$ sharply changes 
from a damped oscillation to an exponential decay 
at some $U_f=U_\ast^{\rm nth}$. 
However, quantitatively the evolution 
of $m$ is very different.
For example, 
the center of the oscillations, their frequency, the damping rate, and the transition point ($U_\ast^{\rm nth}$)
are all different from those of Fig.~2(a) in the main text. This is because the interaction strength that we consider here is already beyond
the one ($U<0.5$) for which the Hartree approximation works. It is thus surprising that the results including higher-order quantum corrections
shown in the main text nevertheless share
the qualitative features 
of the Hartree approximation in this interaction regime.
For $U_i=2.5$ [Fig.~\ref{order parameter Hartree}(b)], the qualitative properties remain
unchanged within the Hartree approximation, 
but the sharpness of the transition is lost after quantum corrections are taken into account by nonequilibrium DMFT [Fig.~2(b) in the main text].

\begin{figure}[t]
\begin{center}
\includegraphics[width=8.5cm]{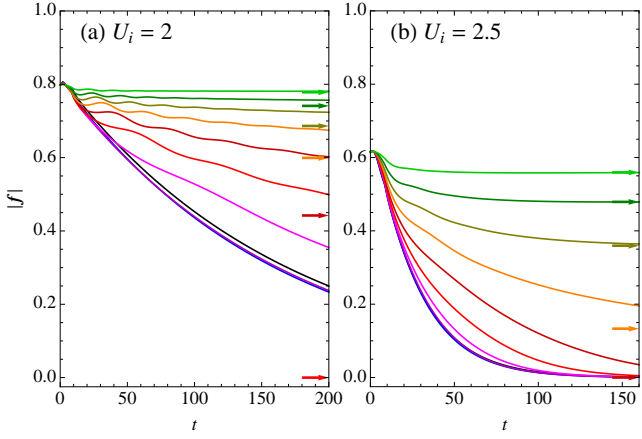}
\caption{Time evolution of $|{\bm f}_{\bm k}|$ at $\epsilon_{\bm k}=0$ calculated with nonequilibrium DMFT
for quenches (a) $U_i=2 \to U_f=1.0,1.1,\dots,1.9$ (from bottom to top)
and (b) $U_i=2.5 \to U_f=1.5, 1.6, \dots, 2.4$ (from bottom to top). The color codes are the same as 
in 
Fig.~2 in the main text. The arrows indicate
the corresponding thermal values reached in the long-time limit.}
\label{fz}
\end{center}
\end{figure}
To see how the Hartree approximation fails for $U>0.5$, 
we plot nonequilibrium DMFT
results for $|{\bm f}_{\bm k}|$ (\ref{length of pseudospin}) at $\epsilon_{\bm k}=0$ (Fermi energy) in Fig.~\ref{fz}.
These quantities would be 
conserved in the Hartree approximation.
At $\epsilon_{\bm k}=0$, $|\bm f_{\bm k}|=|f_{\bm k}^z|$ since the off-diagonal Green functions ($G_{\bm k\sigma}^{AB}, G_{\bm k\sigma}^{BA}$) 
are odd functions of $\epsilon_{\bm k}$. One can calculate $f_{\bm k}^z|_{\epsilon=0}$ from $P_\sigma^a$ [Eq.~(\ref{P}) below]
using the relation $f_{\bm k}^z|_{\epsilon=0}=\frac{1}{2}\sum_\sigma (-i)\sigma(P_\sigma^A-P_\sigma^B)$. 
One can see 
in Fig.~\ref{fz}(a) that $|\bm f_{\bm k}|$ is not conserved even for $U_f=1.0$, 
but starts to decay immediately after the quench without any plateau. 
This suggests that the Hartree equation (\ref{Hartree equation}) is not valid on any time scale, except
in the very weakly correlated regime ($U<0.5$). It was already clear 
from the equilibrium phase diagram [Fig.~1(a)], 
which shows the Hartree phase boundary as a dashed line,
that there are large quantum corrections from higher order diagrams for $U> 0.5$.

Finally, we remark that Eq.~(\ref{Hartree equation}) holds for ``arbitrary filling'', and thus even if the symmetry between
the repulsive and attractive models is not valid any more. This suggests that a nonthermal fixed point similar to
what we have found at half-filling appears also away from half-filling. In fact, we numerically confirmed
with the Hartree equation and the nonequilibrium DMFT
that the slightly doped ($\lesssim 5\%$) system can be trapped in a nonthermal ordered state,
and that the "critical" behavior at the nonthermal fixed point is the same.

\subsection{Momentum distribution function}

In this section, we 
show the derivation of the momentum distribution function
\begin{align}
n_{\bm k}(t)
=
\langle c_{\bm k\sigma}^\dagger(t)c_{\bm k\sigma}(t)\rangle,
\label{nkdef}
\end{align}
and present the numerical results for the slope of the distribution $\partial_\epsilon n_{\bm k}$ at $\epsilon_{\bm k}=0$ (Fermi energy),
which sensitively measures whether and how thermalization takes place.

By definition, one can obtain the momentum distribution (\ref{nkdef}) from the nonequilibrium Green function (\ref{nkabdef}),
\begin{align}
n_{\bm k\sigma}(t)
=
\frac{1}{2}\sum_{ab} (-i)G_{\bm k\sigma}^{ab<}(t,t).
\end{align}
The Green function satisfies the $2\times 2$ Dyson equation (\ref{2x2 Dyson}), which can be reduced to 
a set of $1\times 1$ Dyson equations. To this end, we take the diagonal Green function at $\epsilon_{\bm k}=0$, which we denote by
\begin{align}
P_\sigma^a
\equiv
G_{\bm k\sigma}^{aa}|_{\epsilon=0}
\label{P}
\end{align}
($a=A, B$).
It satisfies
\begin{align}
(i\partial_t+\mu-\Sigma_\sigma^{a})\ast P_\sigma^a
  &=
    \delta_{\mathcal{C}}.
\end{align}
With $P_\sigma^a$, the Green functions at arbitrary $\epsilon_{\bm k}$ are given by
\begin{align}
&(i\partial_t+\mu-\Sigma_\sigma^A-\epsilon_{\bm k}^2P_\sigma^B)\ast G_{\bm k\sigma}^{AA}=\delta_{\mathcal{C}},
\label{GAA}
\\
&(i\partial_t+\mu-\Sigma_\sigma^B-\epsilon_{\bm k}^2P_\sigma^A)\ast G_{\bm k\sigma}^{BB}=\delta_{\mathcal{C}},
\label{GBB}
\\
&G_{\bm k\sigma}^{AB}=\epsilon_{\bm k} G_{\bm k\sigma}^{AA}\ast P_\sigma^B,
\label{GAB}
\\
&G_{\bm k\sigma}^{BA}=\epsilon_{\bm k} G_{\bm k\sigma}^{BB}\ast P_\sigma^A.
\label{GBA}
\end{align}
These are equivalent to solving a set of $1\times 1$ Dyson equations.

\begin{figure}[t]
\begin{center}
\includegraphics[width=8.5cm]{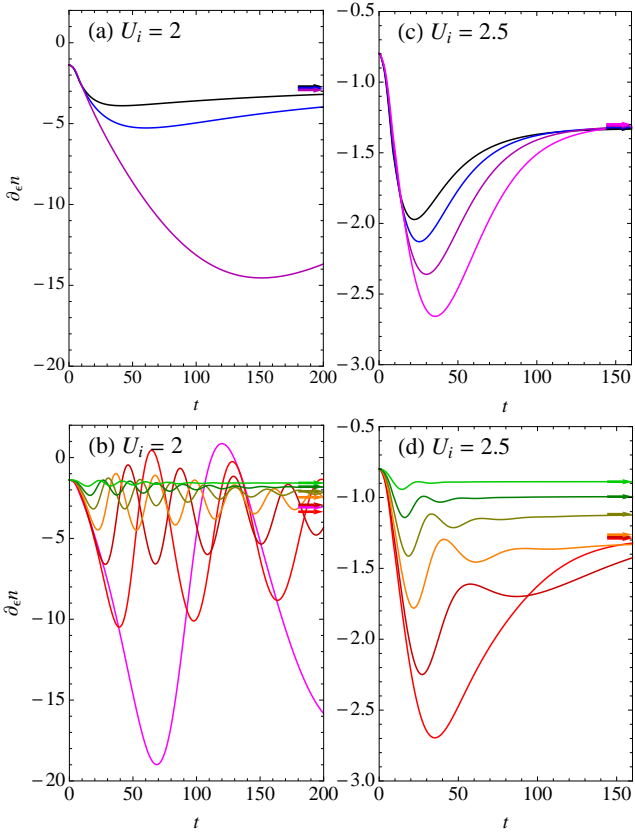}
\caption{(Color online). Time evolution of the slope of the momentum distribution $\partial_\epsilon n_{\bm k}$ at $\epsilon_{\bm k}=0$
calculated with nonequilibrium DMFT
for quenches (a) $U_i=2 \to U_f=1.0, 1.1, 1.2 (<U_\ast^{\rm nth})$ (from top to bottom), 
(b) $U_i=2 \to U_f=1.3, \dots, 1.9 (>U_\ast^{\rm nth})$ (from bottom to top), 
(c) $U_i=2.5 \to U_f=1.5, 1.6, \dots, 1.8 (<U_\ast^{\rm nth})$ (from top to bottom)
and (d) $U_i=2.5 \to U_f=1.9, \dots, 2.4 (>U_\ast^{\rm nth})$ (from bottom to top).
The color codes are the same as Fig.~2 in the main text. The arrows indicate the corresponding thermal values
reached in the long-time limit.}
\label{gradient}
\end{center}
\end{figure}
As is clear from Eqs.~(\ref{GAA})-(\ref{GBA}),
the diagonal Green functions are even functions of $\epsilon_{\bm k}$, 
while the off-diagonal Green functions are odd. Thus we have
\begin{align}
\partial_\epsilon G_{\bm k\sigma}^{AA}|_{\epsilon=0}
  &=
    \partial_\epsilon G_{\bm k\sigma}^{BB}|_{\epsilon=0}
  =
    0.
\end{align}
To get the first derivative of the off-diagonal Green functions, 
we take a derivative with respect to $\epsilon_{\bm k}$ and putting $\epsilon_{\bm k}=0$ in Eq.~(\ref{2x2 Dyson}) to have
\begin{align}
(i\partial_t+\mu-\Sigma_\sigma^a)\ast \partial_\epsilon G_{\bm k\sigma}^{ab}|_{\epsilon=0}
-G_{\bm k\sigma}^{\bar{a}b}|_{\epsilon=0}
  &=
    0.
\end{align}
From this, we obtain
\begin{align}
\partial_\epsilon G_{\bm k\sigma}^{AB}|_{\epsilon=0}
  &=
    P_\sigma^A \ast P_\sigma^B,
\\
\partial_\epsilon G_{\bm k\sigma}^{BA}|_{\epsilon=0}
  &=
    P_\sigma^B \ast P_\sigma^A.
\end{align}
As a result, the slope of the momentum distribution $\partial_\epsilon n_{\bm k\sigma}$ at $\epsilon_{\bm k}=0$ 
is calculated from a convolution,
\begin{align}
\partial_\epsilon n_{\bm k\sigma}(t)|_{\epsilon=0}
  &=
    -\frac{i}{2}(P_\sigma^A \ast P_\sigma^B+P_\sigma^B \ast P_\sigma^A)^<(t,t).
\end{align}

Numerical results for $\partial_\epsilon n$ at $\epsilon_{\bm k}=0$ with the same parameters as Fig.~2 in the main text
are shown in Fig.~\ref{gradient}. For $U_i=2.0$ [Fig.~\ref{gradient}(a), (b)], there are two clearly different behaviors.
When $U_f>1.2$, $\partial_\epsilon n$ coherently oscillates, which is because a wave mode created in the high energy region
of the momentum distribution cascades down to the Fermi energy [Fig.~3(a) in the main text].
The slope $\partial_\epsilon n$ even changes its sign [Fig.~\ref{gradient}(b)] if the amplitude of the oscillations is strong enough.
This never happens in thermal equilibrium.
On longer time scales,
the oscillation slowly damps, and the gradient $\partial_\epsilon n$ finally converges to the corresponding thermal value (thermalization).
A sharp change occurs between $U_f=1.2$ and $1.3$, where the amplitude of the oscillation in $\partial_\epsilon n$ is greatly enhanced, and
even appears to diverge $\propto |U_f-U_\ast^{\rm nth}|^{-1}$ 
with $U_\ast^{\rm nth}=1.23$ 
[see the plot of the inverse of the steepest slope $(\partial_\epsilon n)^{-1}\equiv(\max_t\{|\partial_\epsilon n(t)|\})^{-1}$ 
in Fig.~4(a) of the main text]. As a result, a sharp jump of the momentum distribution starts to appear at the Fermi energy 
[Fig.~3(a) in the main text].
For $U_f\le 1.2$, $\partial_\epsilon n$ overdamps without any oscillation.
This qualitative change of the behavior of $\partial_\epsilon n$
occurs at the same point ($U_f=U_\ast^{\rm nth}$) as that of the order parameter $m$ discussed in the main text.
For $U_i=2.5$ [Fig.~\ref{gradient}(c), (d)],
the enhancement of $\partial_\epsilon n$ is suppressed due to the limited 
life-time 
of the trapped state.
However, one can still see a crossover of the relaxation behavior of $n_{\bm k}$ from damped oscillation to overdamped decay
around $U_f\sim 1.85$,
the value of $U_f$ for which the ``steepest Fermi surface'' is reached during the time evolution. 
Thus we use the steepest $n_{\bm k}$ as a measure of the nonthermal transition point ($U_\ast^{\rm nth}$) in the main text.

\subsection{Quench protocol dependence}

\begin{figure}[t]
\begin{center}
\includegraphics[width=6.5cm]{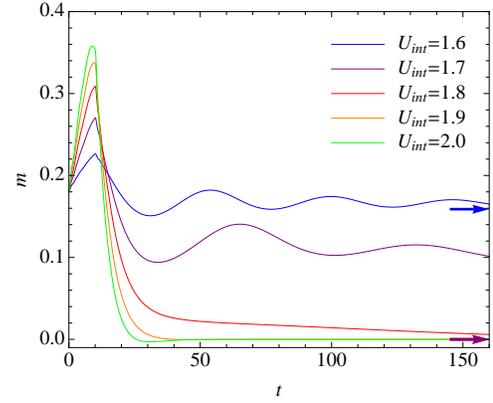}
\caption{(Color online). Time evolution of $m$ obtained from the nonequilibrium DMFT
for a back-and-forth quench $U=U_i\to U_{int}\to U_i$ with the initial $U_i=1.5$, $T=0.056$ and waiting time $t_w=10$.
The arrows indicate the corresponding thermal values reached in the long-time limit.}
\label{quench back and forth}
\end{center}
\end{figure}
In this section, we discuss how the nonthermal quasi-stationary ordered state that has been found in the main text
depends on the interaction quench protocol. In the main text, we concentrated on the linear change
of the interaction, $U(t)=U_i+(U_f-U_i)t/t_q$ ($0\le t\le t_q$) with $t_q=8$ fixed, 
to go across the phase transition boundary. However, it should be noted that the phenomenon is 
not specific to this quench protocol, but is rather general. For example, a qualitatively similar 
nonthermal state is found for a wide range of $t_q$. In particular, we numerically confirmed that a sudden quench (i.e., $t_q=0$) can lead to a 
nonthermal trapped state, although in this case the excitation energy is relatively large 
so that the nonthermal AFM region in Fig.~1(b) of the main text shrinks.

As a further example, we consider a back-and-forth quench protocol, 
in which the interaction changes step-wise as
$U(t)=U_i\, (t<0)$, $=U_{int}\, (0<t<t_w)$, and $=U_i\, (t>t_w)$ with $t_w$ the waiting time (see inset in Fig.~\ref{quench back and forth}).
This back-and-forth quench is a computationally convenient way of injecting energy into the system.  
In Fig.~\ref{quench back and forth}, we show the time evolution of $m$ for $U_i=1.5$ and 
various $U_{int}$ and $t_w=10$. As one can see, the relaxation behavior qualitatively changes 
from a damped oscillation to an exponential decay as $U_{int}$ is varied.
The magnetization $m$ oscillates around a nonzero, nonthermal value for a long time 
even in cases where the thermalized value is zero ($U_{int}=1.7$ in Fig.~\ref{quench back and forth}).
This protocol corresponds to an effective heating without an overall change of the interaction between
the initial and final states. Thus, the trapping phenomenon is not specific to a particular interaction quench,
but can be induced by an external perturbation that increases the effective temperature of the system,
which suggests various possible and realistic ways of reaching a state controlled by the nonthermal fixed point,
such as laser excitations.

\bibliographystyle{apsrev}
\bibliography{dynamical-phase-transition}

\begin{thebibliography}{32}
\expandafter\ifx\csname natexlab\endcsname\relax\def\natexlab#1{#1}\fi
\expandafter\ifx\csname bibnamefont\endcsname\relax
  \def\bibnamefont#1{#1}\fi
\expandafter\ifx\csname bibfnamefont\endcsname\relax
  \def\bibfnamefont#1{#1}\fi
\expandafter\ifx\csname citenamefont\endcsname\relax
  \def\citenamefont#1{#1}\fi
\expandafter\ifx\csname url\endcsname\relax
  \def\url#1{\texttt{#1}}\fi
\expandafter\ifx\csname urlprefix\endcsname\relax\def\urlprefix{URL }\fi
\providecommand{\bibinfo}[2]{#2}
\providecommand{\eprint}[2][]{\url{#2}}

\bibitem[{\citenamefont{Kibble}(1976)}]{Kibble1976}
\bibinfo{author}{\bibfnamefont{T.~W.~B.} \bibnamefont{Kibble}},
  \bibinfo{journal}{J. Phys. A} \textbf{\bibinfo{volume}{9}},
  \bibinfo{pages}{1387} (\bibinfo{year}{1976}).

\bibitem[{\citenamefont{Zurek}(1985)}]{Zurek1985}
\bibinfo{author}{\bibfnamefont{W.~H.} \bibnamefont{Zurek}},
  \bibinfo{journal}{Nature (London)} \textbf{\bibinfo{volume}{317}},
  \bibinfo{pages}{505} (\bibinfo{year}{1985}).

\bibitem[{\citenamefont{Schmitt et~al.}(2008)\citenamefont{Schmitt, Kirchmann,
  Bovensiepen, Moore, Rettig, Krenz, Chu, Ru, Perfetti, Lu
  et~al.}}]{Schmitt2008}
\bibinfo{author}{\bibfnamefont{F.}~\bibnamefont{Schmitt}},
  \bibinfo{author}{\bibfnamefont{P.~S.} \bibnamefont{Kirchmann}},
  \bibinfo{author}{\bibfnamefont{U.}~\bibnamefont{Bovensiepen}},
  \bibinfo{author}{\bibfnamefont{R.~G.} \bibnamefont{Moore}},
  \bibinfo{author}{\bibfnamefont{L.}~\bibnamefont{Rettig}},
  \bibinfo{author}{\bibfnamefont{M.}~\bibnamefont{Krenz}},
  \bibinfo{author}{\bibfnamefont{J.-H.} \bibnamefont{Chu}},
  \bibinfo{author}{\bibfnamefont{N.}~\bibnamefont{Ru}},
  \bibinfo{author}{\bibfnamefont{L.}~\bibnamefont{Perfetti}},
  \bibinfo{author}{\bibfnamefont{D.~H.} \bibnamefont{Lu}},
  \bibnamefont{et~al.}, \bibinfo{journal}{Science}
  \textbf{\bibinfo{volume}{321}}, \bibinfo{pages}{1649} (\bibinfo{year}{2008}).

\bibitem[{\citenamefont{Yusupov et~al.}(2010)\citenamefont{Yusupov, Mertelj,
  Kabanov, Brazovskii, Kusar, Chu, Fisher, and Mihailovic}}]{Yusupov2010}
\bibinfo{author}{\bibfnamefont{R.}~\bibnamefont{Yusupov}},
  \bibinfo{author}{\bibfnamefont{T.}~\bibnamefont{Mertelj}},
  \bibinfo{author}{\bibfnamefont{V.~V.} \bibnamefont{Kabanov}},
  \bibinfo{author}{\bibfnamefont{S.}~\bibnamefont{Brazovskii}},
  \bibinfo{author}{\bibfnamefont{P.}~\bibnamefont{Kusar}},
  \bibinfo{author}{\bibfnamefont{J.-H.} \bibnamefont{Chu}},
  \bibinfo{author}{\bibfnamefont{I.~R.} \bibnamefont{Fisher}},
  \bibnamefont{and}
  \bibinfo{author}{\bibfnamefont{D.}~\bibnamefont{Mihailovic}},
  \bibinfo{journal}{Nat. Phys.} \textbf{\bibinfo{volume}{6}},
  \bibinfo{pages}{681} (\bibinfo{year}{2010}).

\bibitem[{\citenamefont{Fausti et~al.}(2011)\citenamefont{Fausti, Tobey, Dean,
  Kaiser, Dienst, Hoffmann, Pyon, Takayama, Takagi, and
  Cavalleri}}]{Fausti2011}
\bibinfo{author}{\bibfnamefont{D.}~\bibnamefont{Fausti}},
  \bibinfo{author}{\bibfnamefont{R.~I.} \bibnamefont{Tobey}},
  \bibinfo{author}{\bibfnamefont{N.}~\bibnamefont{Dean}},
  \bibinfo{author}{\bibfnamefont{S.}~\bibnamefont{Kaiser}},
  \bibinfo{author}{\bibfnamefont{A.}~\bibnamefont{Dienst}},
  \bibinfo{author}{\bibfnamefont{M.~C.} \bibnamefont{Hoffmann}},
  \bibinfo{author}{\bibfnamefont{S.}~\bibnamefont{Pyon}},
  \bibinfo{author}{\bibfnamefont{T.}~\bibnamefont{Takayama}},
  \bibinfo{author}{\bibfnamefont{H.}~\bibnamefont{Takagi}}, \bibnamefont{and}
  \bibinfo{author}{\bibfnamefont{A.}~\bibnamefont{Cavalleri}},
  \bibinfo{journal}{Science} \textbf{\bibinfo{volume}{331}},
  \bibinfo{pages}{189} (\bibinfo{year}{2011}).

\bibitem[{\citenamefont{Cavalieri et~al.}(2007)\citenamefont{Cavalieri, Muller,
  Uphues, Yakovlev, Baltuska, Horvath, Schmidt, Blumel, Holzwarth, Hendel
  et~al.}}]{Cavalieri2007}
\bibinfo{author}{\bibfnamefont{A.~L.} \bibnamefont{Cavalieri}},
  \bibinfo{author}{\bibfnamefont{N.}~\bibnamefont{Muller}},
  \bibinfo{author}{\bibfnamefont{T.}~\bibnamefont{Uphues}},
  \bibinfo{author}{\bibfnamefont{V.~S.} \bibnamefont{Yakovlev}},
  \bibinfo{author}{\bibfnamefont{A.}~\bibnamefont{Baltuska}},
  \bibinfo{author}{\bibfnamefont{B.}~\bibnamefont{Horvath}},
  \bibinfo{author}{\bibfnamefont{B.}~\bibnamefont{Schmidt}},
  \bibinfo{author}{\bibfnamefont{L.}~\bibnamefont{Blumel}},
  \bibinfo{author}{\bibfnamefont{R.}~\bibnamefont{Holzwarth}},
  \bibinfo{author}{\bibfnamefont{S.}~\bibnamefont{Hendel}},
  \bibnamefont{et~al.}, \bibinfo{journal}{Nature (London)}
  \textbf{\bibinfo{volume}{449}}, \bibinfo{pages}{1029} (\bibinfo{year}{2007}).

\bibitem[{\citenamefont{Bloch et~al.}(2008)\citenamefont{Bloch, Dalibard, and
  Zwerger}}]{BlochDalibardZwerger2008}
\bibinfo{author}{\bibfnamefont{I.}~\bibnamefont{Bloch}},
  \bibinfo{author}{\bibfnamefont{J.}~\bibnamefont{Dalibard}}, \bibnamefont{and}
  \bibinfo{author}{\bibfnamefont{W.}~\bibnamefont{Zwerger}},
  \bibinfo{journal}{Rev. Mod. Phys.} \textbf{\bibinfo{volume}{80}},
  \bibinfo{pages}{885} (\bibinfo{year}{2008}).

\bibitem[{\citenamefont{Berges et~al.}(2004)\citenamefont{Berges, Bors\'anyi,
  and Wetterich}}]{BergesBorsanyiWetterich2004}
\bibinfo{author}{\bibfnamefont{J.}~\bibnamefont{Berges}},
  \bibinfo{author}{\bibfnamefont{S.}~\bibnamefont{Bors\'anyi}},
  \bibnamefont{and}
  \bibinfo{author}{\bibfnamefont{C.}~\bibnamefont{Wetterich}},
  \bibinfo{journal}{Phys. Rev. Lett.} \textbf{\bibinfo{volume}{93}},
  \bibinfo{pages}{142002} (\bibinfo{year}{2004}).

\bibitem[{\citenamefont{Moeckel and Kehrein}(2008)}]{MoeckelKehrein2008}
\bibinfo{author}{\bibfnamefont{M.}~\bibnamefont{Moeckel}} \bibnamefont{and}
  \bibinfo{author}{\bibfnamefont{S.}~\bibnamefont{Kehrein}},
  \bibinfo{journal}{Phys. Rev. Lett.} \textbf{\bibinfo{volume}{100}},
  \bibinfo{pages}{175702} (\bibinfo{year}{2008}).

\bibitem[{\citenamefont{Eckstein et~al.}(2009)\citenamefont{Eckstein, Kollar,
  and Werner}}]{EcksteinKollarWerner2009}
\bibinfo{author}{\bibfnamefont{M.}~\bibnamefont{Eckstein}},
  \bibinfo{author}{\bibfnamefont{M.}~\bibnamefont{Kollar}}, \bibnamefont{and}
  \bibinfo{author}{\bibfnamefont{P.}~\bibnamefont{Werner}},
  \bibinfo{journal}{Phys. Rev. Lett.} \textbf{\bibinfo{volume}{103}},
  \bibinfo{pages}{056403} (\bibinfo{year}{2009}).

\bibitem[{\citenamefont{Kollar et~al.}(2011)\citenamefont{Kollar, Wolf, and
  Eckstein}}]{KollarWolfEckstein2011}
\bibinfo{author}{\bibfnamefont{M.}~\bibnamefont{Kollar}},
  \bibinfo{author}{\bibfnamefont{F.~A.} \bibnamefont{Wolf}}, \bibnamefont{and}
  \bibinfo{author}{\bibfnamefont{M.}~\bibnamefont{Eckstein}},
  \bibinfo{journal}{Phys. Rev. B} \textbf{\bibinfo{volume}{84}},
  \bibinfo{pages}{054304} (\bibinfo{year}{2011}).

\bibitem[{\citenamefont{Berges et~al.}(2008)\citenamefont{Berges, Rothkopf, and
  Schmidt}}]{BergesRothkoptSchmidt2008}
\bibinfo{author}{\bibfnamefont{J.}~\bibnamefont{Berges}},
  \bibinfo{author}{\bibfnamefont{A.}~\bibnamefont{Rothkopf}}, \bibnamefont{and}
  \bibinfo{author}{\bibfnamefont{J.}~\bibnamefont{Schmidt}},
  \bibinfo{journal}{Phys. Rev. Lett.} \textbf{\bibinfo{volume}{101}},
  \bibinfo{pages}{041603} (\bibinfo{year}{2008}).

\bibitem[{\citenamefont{Hohenberg and Halperin}(1977)}]{HohenbergHalperin1977}
\bibinfo{author}{\bibfnamefont{P.~C.} \bibnamefont{Hohenberg}}
  \bibnamefont{and} \bibinfo{author}{\bibfnamefont{B.~I.}
  \bibnamefont{Halperin}}, \bibinfo{journal}{Rev. Mod. Phys.}
  \textbf{\bibinfo{volume}{49}}, \bibinfo{pages}{435} (\bibinfo{year}{1977}).

\bibitem[{\citenamefont{Polkovnikov et~al.}(2011)\citenamefont{Polkovnikov,
  Sengupta, Silva, and
  Vengalattore}}]{PolkovnikovSenguptaSilvaVengalattore2011}
\bibinfo{author}{\bibfnamefont{A.}~\bibnamefont{Polkovnikov}},
  \bibinfo{author}{\bibfnamefont{K.}~\bibnamefont{Sengupta}},
  \bibinfo{author}{\bibfnamefont{A.}~\bibnamefont{Silva}}, \bibnamefont{and}
  \bibinfo{author}{\bibfnamefont{M.}~\bibnamefont{Vengalattore}},
  \bibinfo{journal}{Rev. Mod. Phys.} \textbf{\bibinfo{volume}{83}},
  \bibinfo{pages}{863} (\bibinfo{year}{2011}).

\bibitem[{\citenamefont{Feldman}(2005)}]{Feldman2005}
\bibinfo{author}{\bibfnamefont{D.~E.} \bibnamefont{Feldman}},
  \bibinfo{journal}{Phys. Rev. Lett.} \textbf{\bibinfo{volume}{95}},
  \bibinfo{pages}{177201} (\bibinfo{year}{2005}).

\bibitem[{\citenamefont{Mitra et~al.}(2006)\citenamefont{Mitra, Takei, Kim, and
  Millis}}]{MitraTakeiBaekMillis2006}
\bibinfo{author}{\bibfnamefont{A.}~\bibnamefont{Mitra}},
  \bibinfo{author}{\bibfnamefont{S.}~\bibnamefont{Takei}},
  \bibinfo{author}{\bibfnamefont{Y.~B.} \bibnamefont{Kim}}, \bibnamefont{and}
  \bibinfo{author}{\bibfnamefont{A.~J.} \bibnamefont{Millis}},
  \bibinfo{journal}{Phys. Rev. Lett.} \textbf{\bibinfo{volume}{97}},
  \bibinfo{pages}{236808} (\bibinfo{year}{2006}).

\bibitem[{\citenamefont{Kollath et~al.}(2007)\citenamefont{Kollath, L\"auchli,
  and Altman}}]{KollathLauchliAltman2007}
\bibinfo{author}{\bibfnamefont{C.}~\bibnamefont{Kollath}},
  \bibinfo{author}{\bibfnamefont{A.~M.} \bibnamefont{L\"auchli}},
  \bibnamefont{and} \bibinfo{author}{\bibfnamefont{E.}~\bibnamefont{Altman}},
  \bibinfo{journal}{Phys. Rev. Lett.} \textbf{\bibinfo{volume}{98}},
  \bibinfo{pages}{180601} (\bibinfo{year}{2007}).

\bibitem[{\citenamefont{Barmettler et~al.}(2009)\citenamefont{Barmettler, Punk,
  Gritsev, Demler, and Altman}}]{BarmettlerPunkGritsevDemlerAltman2009}
\bibinfo{author}{\bibfnamefont{P.}~\bibnamefont{Barmettler}},
  \bibinfo{author}{\bibfnamefont{M.}~\bibnamefont{Punk}},
  \bibinfo{author}{\bibfnamefont{V.}~\bibnamefont{Gritsev}},
  \bibinfo{author}{\bibfnamefont{E.}~\bibnamefont{Demler}}, \bibnamefont{and}
  \bibinfo{author}{\bibfnamefont{E.}~\bibnamefont{Altman}},
  \bibinfo{journal}{Phys. Rev. Lett.} \textbf{\bibinfo{volume}{102}},
  \bibinfo{pages}{130603} (\bibinfo{year}{2009}).

\bibitem[{\citenamefont{Koga and Werner}(2011)}]{KogaWerner2011}
\bibinfo{author}{\bibfnamefont{A.}~\bibnamefont{Koga}} \bibnamefont{and}
  \bibinfo{author}{\bibfnamefont{P.}~\bibnamefont{Werner}},
  \bibinfo{journal}{Phys. Rev. A} \textbf{\bibinfo{volume}{84}},
  \bibinfo{pages}{023638} (\bibinfo{year}{2011}).

\bibitem[{\citenamefont{Werner et~al.}(2012)\citenamefont{Werner, Tsuji, and
  Eckstein}}]{WernerTsujiEckstein2012}
\bibinfo{author}{\bibfnamefont{P.}~\bibnamefont{Werner}},
  \bibinfo{author}{\bibfnamefont{N.}~\bibnamefont{Tsuji}}, \bibnamefont{and}
  \bibinfo{author}{\bibfnamefont{M.}~\bibnamefont{Eckstein}},
  \bibinfo{journal}{Phys. Rev. B} \textbf{\bibinfo{volume}{86}},
  \bibinfo{pages}{205101} (\bibinfo{year}{2012}).

\bibitem[{\citenamefont{Tsuji et~al.}(2011)\citenamefont{Tsuji, Oka, Werner,
  and Aoki}}]{TsujiOkaWernerAoki2011}
\bibinfo{author}{\bibfnamefont{N.}~\bibnamefont{Tsuji}},
  \bibinfo{author}{\bibfnamefont{T.}~\bibnamefont{Oka}},
  \bibinfo{author}{\bibfnamefont{P.}~\bibnamefont{Werner}}, \bibnamefont{and}
  \bibinfo{author}{\bibfnamefont{H.}~\bibnamefont{Aoki}},
  \bibinfo{journal}{Phys. Rev. Lett.} \textbf{\bibinfo{volume}{106}},
  \bibinfo{pages}{236401} (\bibinfo{year}{2011}).

\bibitem[{\citenamefont{Tsuji et~al.}(2012)\citenamefont{Tsuji, Oka, Aoki, and
  Werner}}]{TsujiOkaAokiWerner2012}
\bibinfo{author}{\bibfnamefont{N.}~\bibnamefont{Tsuji}},
  \bibinfo{author}{\bibfnamefont{T.}~\bibnamefont{Oka}},
  \bibinfo{author}{\bibfnamefont{H.}~\bibnamefont{Aoki}}, \bibnamefont{and}
  \bibinfo{author}{\bibfnamefont{P.}~\bibnamefont{Werner}},
  \bibinfo{journal}{Phys. Rev. B} \textbf{\bibinfo{volume}{85}},
  \bibinfo{pages}{155124} (\bibinfo{year}{2012}).

\bibitem[{\citenamefont{Georges et~al.}(1996)\citenamefont{Georges, Kotliar,
  Krauth, and Rozenberg}}]{GeorgesKotliarKrauthRozenberg1996}
\bibinfo{author}{\bibfnamefont{A.}~\bibnamefont{Georges}},
  \bibinfo{author}{\bibfnamefont{G.}~\bibnamefont{Kotliar}},
  \bibinfo{author}{\bibfnamefont{W.}~\bibnamefont{Krauth}}, \bibnamefont{and}
  \bibinfo{author}{\bibfnamefont{M.~J.} \bibnamefont{Rozenberg}},
  \bibinfo{journal}{Rev. Mod. Phys.} \textbf{\bibinfo{volume}{68}},
  \bibinfo{pages}{13} (\bibinfo{year}{1996}).

\bibitem[{\citenamefont{Freericks et~al.}(2006)\citenamefont{Freericks,
  Turkowski, and Zlati\ifmmode~\acute{c}\else
  \'{c}\fi{}}}]{FreericksTurkowskiZlatic2006}
\bibinfo{author}{\bibfnamefont{J.~K.} \bibnamefont{Freericks}},
  \bibinfo{author}{\bibfnamefont{V.~M.} \bibnamefont{Turkowski}},
  \bibnamefont{and}
  \bibinfo{author}{\bibfnamefont{V.}~\bibnamefont{Zlati\ifmmode~\acute{c}\else
  \'{c}\fi{}}}, \bibinfo{journal}{Phys. Rev. Lett.}
  \textbf{\bibinfo{volume}{97}}, \bibinfo{pages}{266408}
  (\bibinfo{year}{2006}).

\bibitem[{\citenamefont{Metzner and Vollhardt}(1989)}]{MetznerVollhardt1989}
\bibinfo{author}{\bibfnamefont{W.}~\bibnamefont{Metzner}} \bibnamefont{and}
  \bibinfo{author}{\bibfnamefont{D.}~\bibnamefont{Vollhardt}},
  \bibinfo{journal}{Phys. Rev. Lett.} \textbf{\bibinfo{volume}{62}},
  \bibinfo{pages}{324} (\bibinfo{year}{1989}).

\bibitem[{\citenamefont{Werner et~al.}(2005)\citenamefont{Werner, Parcollet,
  Georges, and Hassan}}]{WernerParcolletGeorgesHassan2005}
\bibinfo{author}{\bibfnamefont{F.}~\bibnamefont{Werner}},
  \bibinfo{author}{\bibfnamefont{O.}~\bibnamefont{Parcollet}},
  \bibinfo{author}{\bibfnamefont{A.}~\bibnamefont{Georges}}, \bibnamefont{and}
  \bibinfo{author}{\bibfnamefont{S.~R.} \bibnamefont{Hassan}},
  \bibinfo{journal}{Phys. Rev. Lett.} \textbf{\bibinfo{volume}{95}},
  \bibinfo{pages}{056401} (\bibinfo{year}{2005}).

\bibitem[{Sup()}]{Supplementary}
\bibinfo{note}{See Supplemental Material for the Hartree approximation, the
  momentum distribution function, and the quench protocol dependence.}

\bibitem[{\citenamefont{Barankov and Levitov}(2006)}]{BarankovLevitov2006}
\bibinfo{author}{\bibfnamefont{R.~A.} \bibnamefont{Barankov}} \bibnamefont{and}
  \bibinfo{author}{\bibfnamefont{L.~S.} \bibnamefont{Levitov}},
  \bibinfo{journal}{Phys. Rev. Lett.} \textbf{\bibinfo{volume}{96}},
  \bibinfo{pages}{230403} (\bibinfo{year}{2006}).

\bibitem[{\citenamefont{Yuzbashyan and Dzero}(2006)}]{YuzbashyanDzero2006}
\bibinfo{author}{\bibfnamefont{E.~A.} \bibnamefont{Yuzbashyan}}
  \bibnamefont{and} \bibinfo{author}{\bibfnamefont{M.}~\bibnamefont{Dzero}},
  \bibinfo{journal}{Phys. Rev. Lett.} \textbf{\bibinfo{volume}{96}},
  \bibinfo{pages}{230404} (\bibinfo{year}{2006}).

\bibitem[{lim()}]{limitation}
\bibinfo{note}{In the previous case ($U_i=2$), the crossover from dephasing to
  thermalization could not be observed, because it occurs at times which cannot
  be reached in our simulation.}

\bibitem[{\citenamefont{Shiba}(1972)}]{Shiba1972}
\bibinfo{author}{\bibfnamefont{H.}~\bibnamefont{Shiba}},
  \bibinfo{journal}{Prog. Theor. Phys.} \textbf{\bibinfo{volume}{48}},
  \bibinfo{pages}{2171} (\bibinfo{year}{1972}).

\bibitem[{\citenamefont{Anderson}(1958)}]{Anderson1958}
\bibinfo{author}{\bibfnamefont{P.~W.} \bibnamefont{Anderson}},
  \bibinfo{journal}{Phys. Rev.} \textbf{\bibinfo{volume}{112}},
  \bibinfo{pages}{1900} (\bibinfo{year}{1958}).

\end{thebibliography}

\end{document}